# Enrichment of Armchair Carbon Nanotubes via Density Gradient Ultracentrifugation: Raman Spectroscopy Evidence

Erik H. Hároz[†], William D. Rice[†], Benjamin Y. Lu[†], Saunab Ghosh[‡], Robert H. Hauge[‡], R. Bruce Weisman[‡], Stephen K. Doorn[§*], and Junichiro Kono[†*]

*Department of Electrical and Computer Engineering, Rice University, Houston, Texas 77005, USA; Department of Chemistry, Rice University, Houston, Texas 77005, USA; Chemistry Division, Los Alamos National Laboratory, Los Alamos, New Mexico 87545, USA*

AUTHOR EMAIL ADDRESS: kono@rice.edu

**RECEIVED DATE (to be automatically inserted after your manuscript is accepted if required according to the journal that you are submitting your paper to)**

CORRESPONDING AUTHOR FOOTNOTE

* To whom correspondence should be addressed.  E-mail: kono@rice.edu, skdoorn@lanl.gov.

†Department of Electrical and Computer Engineering, Rice University.

‡Department of Chemistry, Rice University.

§Chemistry Division, Los Alamos National Laboratory.

ABSTRACT

We have used resonant Raman scattering spectroscopy to fully analyze the relative abundances of different ($n,m$) species in single-walled carbon nanotube samples that are metallically enriched by density gradient ultracentrifugation. Strikingly, the data clearly show that our density gradient ultracentrifugation process enriches the metallic fractions in armchair and near-armchair species. We observe that armchair carbon nanotubes constitute more than 50% of each ($2n + m$) family.

MANUSCRIPT TEXT

As exceptional one-dimensional conductors, metallic single-walled carbon nanotubes are ideal candidates for a variety of electronic applications[1,2] such as nanocircuit components and power transmission cables. In particular, ($n,n$)-chirality, or 'armchair', metallic nanotubes are predicted to be truly gapless and intrinsically insensitive to disorder,[3,4] consistent with experimentally observed ballistic conduction behavior[5,6]. Unfortunately, progress towards such applications has been slowed by the inherent problem of nanotube synthesis whereby both semiconducting and metallic nanotubes are produced. Here, we use a metallic nanotube enrichment process based on a modified approach to the density gradient ultracentrifugation (DGU) technique first introduced by Arnold and coworkers[7,8] to create metallic-enriched nanotube samples. Using resonant Raman scattering spectroscopy, we have fully analyzed the relative abundances of the metallic species present in the sample both before and after DGU. Strikingly, the data clearly show that our DGU process enriches the metallic fractions in armchair and near-armchair species.

The DGU technique can separate as-produced nanotubes, suspended in aqueous solution by multiple surfactants and/or salts, based on conduction type, producing samples comprised almost entirely of metallic or semiconducting nanotubes[7-10]. Typically, type purity has been assessed by a combination of photoluminescence excitation (PLE) spectroscopy and optical absorption spectroscopy[7-9,11] or by electrical conductivity measurements[8,9,12-14]. While PLE is a powerful approach, it can only be used to investigate semiconducting nanotubes. Moreover, broad overlapping peaks in absorption spectra and a lack of structural sensitivity in conductivity measurements make these techniques unsuitable for resolving the precise (*n,m*) makeup of metallic fractions. As a result, both absorption and conductivity measurements provide little information on the physical mechanism responsible for type-dependent enrichments.

Currently, resonant Raman scattering (RRS) spectroscopy is the only optical route to unambiguous assignments of metallic features to specific (*n,m*) species due to its ability to identify and correlate diameter-dependent phonons with chirality-specific optical transitions for all SWNTs. We present results of RRS measurements on DGU metallic-enriched fractions performed over a broad range of excitation wavelengths (440-850 nm) such that all (*n,m*) species remaining in the enriched material were identified. When compared to as-produced single-walled carbon nanotubes (AP-SWNT) material, data from a metallic-enriched SWNT (ME-SWNT) sample indicate almost complete suppression of semiconducting SWNTs, confirming observations by PLE and absorption. Surprisingly, a comparison between the resonant Raman excitation profiles (REPs) of armchair SWNTs and the other metallic species indicates a *strong enrichment of species of large chiral angle (>19º) and, in particular, armchair SWNTs*. By combining results from such data, we have gained valuable insight into the phenomenological nature of DGU-based type separation.

We followed the separation approach given by Yanagi *et al.*[9] with only small variations[15]. Figure 1a is a photograph of typical AP- and ME-SWNT materials, suspended in aqueous media using surfactant, whose absorption spectra are displayed in Fig. 1b. The AP-SWNT spectrum shows well-defined peaks corresponding to the (n,m)-dependent, first, second, and third semiconducting [$E_{11}^S$ (870-1600 nm), $E_{22}^S$ (550-870 nm), and $E_{33}^S$ (UV-490 nm)] and first metallic [$E_{11}^M$ (440-670 nm)] optical transitions typical for individualized HiPco SWNTs[16]. In contrast, the ME-SWNT spectrum shows no features in the $E_{11}^S$ and $E_{22}^S$ regions, indicative of the absence of semiconducting SWNTs. Also apparent is the increased peak-to-valley ratio in the $E_{11}^M$ region of the ME-SWNTs, demonstrating a higher degree of individuality of the metallic chiralities and removal of the overlap between semiconducting and metallic peaks. An assessment of type purity can be determined by integrating the peak areas in the $E_{11}^M$ and $E_{22}^S$ regions[8-9,11]. This results in a metallic purity of ~ 98% for the ME-SWNT sample and 40% for the AP-SWNT sample. While such values are useful for determining the overall success of metallic enrichment, unknown baselines and overlapping absorption features make this technique unreliable.

The specific (*n,m*) semiconducting species remaining in the ME-SWNT sample can be determined via PLE[16-18]. Figures 1c and 1d show PLE data for the AP-SWNT and ME-SWNT samples, respectively. While bright emission features emanate from the former (c), no emission was detected from the latter (d), suggesting the absence of semiconducting SWNTs in ME-SWNT. One alternative explanation for the lack of emission is the extensive bundling of semiconducting nanotubes containing at least one metallic member, which would quench PL. However, the well-defined absorption features in the $E_{11}^M$ region and the lack of any features in the $E_{11}^S$ and $E_{22}^S$ regions in Fig. 1b exclude this possibility. Consequently, we can state that the ME-SWNT fraction consists almost entirely of metallic SWNT species (see supplementary material for further discussion of semiconductors[15]).

RRS spectroscopy possesses the advantage of being able to detect all (*n,m*) species present, regardless of electronic type or aggregation state. By combining spectra obtained using a variety of excitation sources, all (*n,m*) species can be identified by correlating the resonance of the excitation wavelength to specific optical transitions with the diameter-dependent radial breathing mode (RBM) frequency[19-21]. Figures 2a and 2b plot RRS intensity for the AP-SWNT and ME-SWNT samples, respectively, as a function of excitation wavelength (562-670 nm) and Raman shift. For HiPco samples in this excitation range, RRS occurs for metallic and semiconducting SWNTs with diameters in the 0.95-1.36 nm and 0.68-0.90 nm ranges via $E_{11}^M$ and $E_{22}^S$, respectively. In the ME-SWNT sample (Fig. 2b), RBMs from small-diameter semiconductors [e.g., (11,1), (7,5), (7,6), and (8,3)] are almost completely suppressed with the dominant contribution coming from metallic nanotubes of the (2*n* + *m*) = 27 and 30 families.

Further examination of the cluster of peaks of family 27 (insets for Figs. 2a and 2b) reveals an unexpected change in the relative Raman intensities of the (n,m) members of family 27 between AP-SWNT and ME-SWNT. The strong peaks for the small-chiral-angle species [(11,5) and (12,3)] in AP-SWNT decrease in intensity through DGU, relative to the armchair (9,9) and near-armchair (10,7) species (see Fig. 2b inset). At higher photon energies, this trend becomes even more selective. Figures 2c (AP-SWNT) and 2d (ME-SWNT) show RRS at 445-500 nm, where resonances occur for small-diameter metallic nanotubes (0.72-0.95 nm) and larger-diameter semiconducting (0.97-1.36 nm) nanotubes via $E_{11}^M$ and $E_{33}^S$, respectively. Again, strong suppression of semiconducting RBMs is clear, as well as suppression of some metallic RBMs. Closer examination of Figs. 2c and 2d reveals that metallic-enrichment of families 18 and 21 is mostly due to armchair nanotubes, with only the (7,7) and (6,6) species remaining in ME-SWNT. Suppression occurred for the (8,5) and (9,3) species of family 21 and the (7,4) and (8,2) species of family 18 [(9,3) was observed in AP-SWNTs via 514.5 nm discrete excitation[15]].

To quantify this chiral-angle-dependent metallic enrichment, we constructed REPs from the data contained in Fig. 2. Figures 3a and 3b show REPs for family 27 for AP-SWNT and ME-SWNT, respectively. The Raman intensity, $I_{Raman}$, for a particular RBM [i.e., (n,m) species] can be written as a function of excitation energy $E_{laser}$ such that

$$I_{Raman} = gN \left| \sum_{i,j} \frac{M_{e-o}^{g,i} M_{e-ph}^{g,i} M_{e-o}^{i,j}}{(E_{laser} - E_{ii} - i\gamma)(E_{laser} - E_{ii} - \hbar\omega_{ph} - i\gamma)} \right|^2 \quad (1)$$

where $g$ is an experimental pre-factor, $N$ is the relative population of the (n,m) species probed, $E_{ii}$ is the optical transition energy, $\gamma$ is the electronic broadening factor, $\hbar\omega_{ph}$ is the phonon energy, $M_{e-ph}$ is the exciton-phonon coupling matrix element, $M_{e-o}$ is the exciton-photon coupling matrix element, and the summation is over electronic states[22]. We used Equation (1) to analyze the REP data in Fig. 3 to determine the relative (n,m) populations for families 27 and 30 for both AP-SWNT and ME-SWNT, and the results are summarized in Table 1. Smaller-chiral-angle species show significant suppression, with the (12,3) registering no signal after DGU. In contrast, the armchair and near-armchair chiralities become the dominant elements, rising together from a relative population of 70% to 98% in family 27; analogous but less dramatic behavior is seen with family 30 (see Supplementary Material for further details on population analysis). The noticeably high relative armchair population, even in the AP-SWNT sample, is surprising but consistent with recent theoretical work suggesting that armchair nanotubes should be the most populous species in the distribution of nanotubes synthesized by several different methods[23]. Experimentally, this is difficult to verify, since the armchair species have a weak Raman response due to their small exciton-phonon coupling and, as a result, are obscured by semiconducting tubes and other metallic tubes[19,20,22,24,25].

A similar chiral trend in DGU-based, metallic enrichment has been previously reported by Sato *et al.*[27], using transmission electron microscopy nanodiffraction measurements. However, the nature of this approach yields only a small statistical sampling of the entire sample. By employing RRS, we extracted information on nanotube ensembles with data that is an average over a macroscopic population of each (*n,m*) species. One important advantage of this is the ability to take an in-depth look at armchair nanotubes as a macroscopic ensemble. Several previous studies[19,20,22,24,25] have noted the exceedingly small Raman signal for armchairs as compared to other metallic species, even of similar diameter, with some theoretical estimates showing that armchairs possess exciton-phonon couplings an order of magnitude smaller than zigzag and near-zigzag species of the same family[22,25,26]. Despite this, armchairs (6,6) through (11,11) are clearly displayed in ME-SWNT, as shown in Figs. 4a and 4b. Of particular note is the clarity with which the armchairs may be observed in Fig. 4b as compared to Fig. 4a.

Taken together, we have clear evidence that in DGU armchair and large-chiral-angle (> 19º) species are enriched, while zigzag and near-zigzag metallic species and nearly all semiconductors are removed. One can hypothesize that such chiral angle selectivity might stem from a specificity in the nanotube interaction with one of the chiral surfactants, sodium cholate and/or sodium deoxycholate, which has been suggested by Green *et al.*[28] for enantiomer enrichment of left- and right-handed (6,5) nanotubes. While the achiral surfactant sodium dodecyl sulfate (SDS) has been shown to be necessary for metallic enrichment[8-10] [the DGU-based enrichments of Niyogi *et al.*[10], which employ SDS and alkali salts (both achiral compounds), demonstrate enrichment of metallic species without any chiral angle bias], it may be the ease of stacking of the chiral surfactant onto the SWNT surface due to registry-matching and sterics between the two that leads to the chiral angle selectivity[29,30].

ACKNOWLEDGMENTS

This work was supported by the DOE/BES through Grant No. DEFG02-06ER46308, the Robert A. Welch Foundation through Grant Nos. C-1509 and C-0807, the Air Force Research Laboratories under contract number FA8650-05-D-5807, the NSF through Grant No. CHE-0809020, and the LANL LDRD Program. We would like to thank Kazuhiro Yanagi, Carter Kittrell, Wade Adams, Noe Alvarez, and Cary Pint for useful and stimulating discussions.

REFERENCES

1. Dekker, C. *Physics Today* **1999**, *52* (5), 22-28.

2. Baughman, R. H.; Zakhidov, A. A.; de Heer, W. A. *Science* **2002**, *297*, 787-792.

3. Ando, T.; Nakanishi, T. *J. Phys. Soc. Jpn.* **1998**, *67*, 1704-1713.

4. White, C. T.; Todorov, T. N. *Nature* **1998**, *393*, 240-242.

5. Tans, S. J.; Devoret, M. H.; Dai, H.; Thess, A.; Smalley, R. E.; Geerligs, L. J.; Dekker, C. *Nature* **1997**, *386*, 474-477.

6. Javey, A.; Guo, J.; Wang, Q.; Lundstrom, M.; Dai, H. *Nature* **2003**, *424*, 654-657.

7. Arnold, M. S.; Stupp, S. I.; Hersam, M. C. *Nano Lett.* **2005**, *5*, 713-718.

8. Arnold, M. S.; Green, A. A.; Hulvat, J. F.; Stupp, S. I.; Hersam, M. C. *Nature Nanotech.* **2006**, *1*, 60-65.

9. Yanagi, K.; Miyata, Y.; Kataura, H. *Appl. Phys. Exp.* **2008**, *1*, 034003.

10. Niyogi, S.; Densmore, C. G.; Doorn, S. K. *J. Am. Chem. Soc.* **2009**, *131*, 1144-1153.

11. Miyata, Y.; Yanagi, K.; Maniwa, Y.; Kataura, H. *J. Phys. Chem. C* **2008**, *112*, 13187-13191.

12. Blackburn, J. L.; Barnes, T. M.; Beard, M. C.; Kim, Y. H.; Tenent, R. C.; McDonald, T. J.; To, B.; Coutts, T. J.; Heben, M. J. *ACS Nano* **2008**, *2*, 1266-1274.

13. Green, A. A.; Hersam, M. C. *Nano Lett.* **2008**, *8*, 1417-1422.

14. Miyata, Y.; Yanagi, K.; Maniwa, Y.; Kataura, H. *J. Phys. Chem. C* **2008**, *112*, 3591-3596.

15. See the online Supplementary Information for details.

16. O'Connell, M. J.; Bachilo, S. M.; Huffman, C. B.; Moore, V. M.; Strano, M. S.; Haroz, E. H.; Rialon, K. L.; Boul, P. J.; Noon, W. H.; Kittrell, C.; Ma, J.; Hauge, R. H.; Weisman, R. B.; Smalley, R. E. *Science* **2002**, *297*, 593-596.

17. Bachilo, S. M.; Strano, M. S.; Kittrell, C.; Hauge, R. H.; Smalley, R. E.; Weisman, R. B. *Science* **2002**, *298*, 2361-2366.

18. Weisman, R. B.; Bachilo, S. M. *Nano Lett.* **2003**, *3*, 1235-1238.

19. Maultzsch, J.; Telg, H.; Reich, S.; Thomsen, C. *Phys. Rev. B* **2005**, *72*, 205438.

20. Fantini, C.; Jorio, A.; Souza, M.; Strano, M. S.; Dresselhaus, M. S.; Pimenta, M. A. *Phys. Rev. Lett.* **2004**, *93*, 147406.

21. Doorn, S. K.; Heller, D. A.; Barone, P. W.; Ursey, M. L.; Strano, M. S. *Appl. Phys. A* **2004**, *78*, 1147-1155.

22. Jiang, J.; Saito, R.; Sato, K.; Park, J. S.; Samsonidze, Ge. G.; Jorio, A.; Dresselhaus, G.; Dresselhaus, M. S. *Phys. Rev. B* **2007**, *75*, 035405.

23. Ding, F.; Harutyunyan, A.; Yakobson, B. I. *PNAS* **2009**, *106*, 2506-2509.

24. Strano, M. S.; Doorn, S. K.; Haroz, E. H.; Kittrell, C.; Hauge, R. H.; Smalley, R. E. *Nano Lett.* **2003**, *3*, 1091-1096.

25. Machon, M.; Reich, S.; Telg, H.; Maultzsch, J.; Ordejón, P.; Thomsen, C. *Phys. Rev. B*. **2005**, *71*, 035416.

26. Goupalov, S. V.; Satishkumar, B. C.; Doorn, S. K. *Phys. Rev. B* **2006**, *73*, 115401.

27. Sato, Y.; Yanagi, K.; Miyata, Y.; Suenaga, K.; Kataura, H.; Iijima, S. *Nano Lett.* **2008**, *8*, 3151-3154.

28. Green, A. A.; Duch, M. C.; Hersam, M. C. *Nano Res.* **2009**, *2*, 69-77.

29. Mukhopadhyay, S.; Maitra, U. *Curr. Sci.* **2004**, *87*, 1666-1683.

30. Arnold, M. S.; Suntivich, J.; Stupp, S. I.; Hersam, M. C. *ACS Nano* **2008**, *2*, 2291-2300.

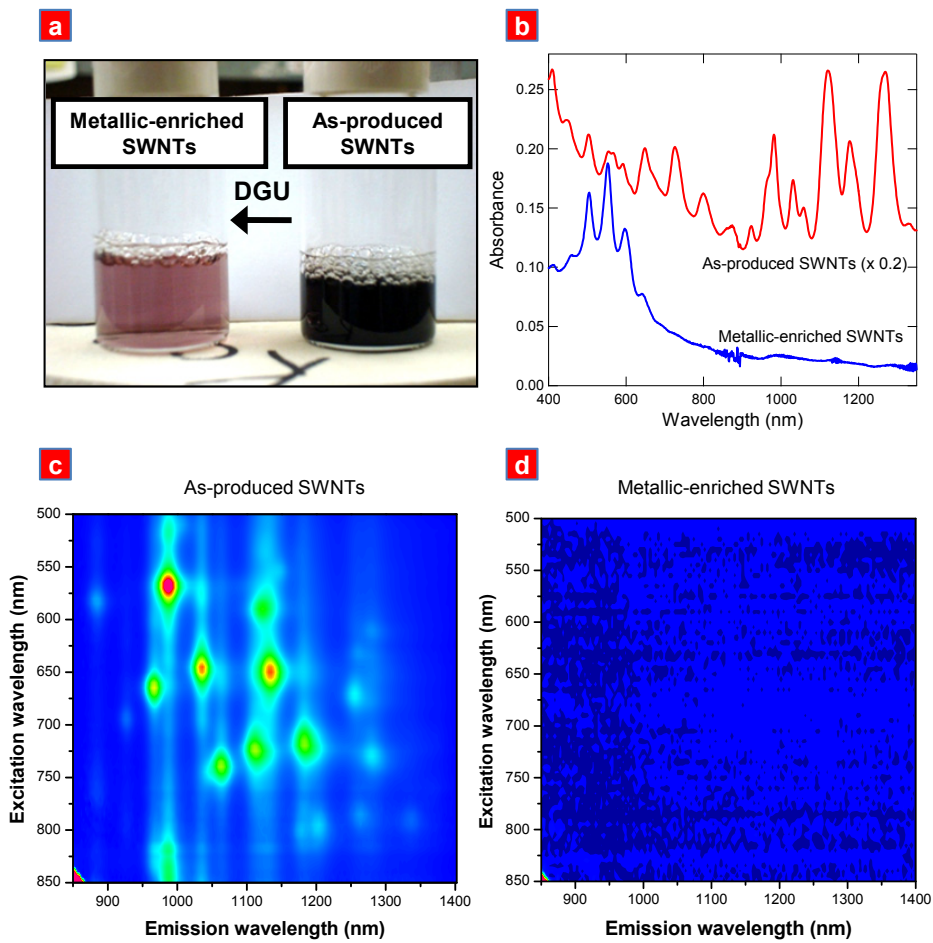

**Figure 1.** Comparison between as-produced and metallic-enriched single-walled carbon nanotube suspensions. (a) A typical metallic type-enrichment (left) of HiPco SWNTs using density gradient ultracentrifugation compared to the as-produced HiPco material (right), both suspended in aqueous solution using surfactants. (b) Absorption spectra of metallic-enriched HiPco material and as-produced HiPco material. The as-produced material has been scaled by 0.2 for ease of comparison to the enriched material. A clear suppression of semiconducting absorption features (650-1300 nm) is evident as well as a sharpening of absorption features due to metallic nanotubes (450-650 nm) in the metallic-enriched SWNT sample as compared to as-produced SWNTs. (c,d) Photoluminescence excitation maps of as-produced HiPco (c) and metallic-enriched (d) HiPco materials. Signal in (d) has been scaled by 10 in photoluminescence intensity to show any possible weak features and yet no signal could be observed from the metallic-enriched sample.

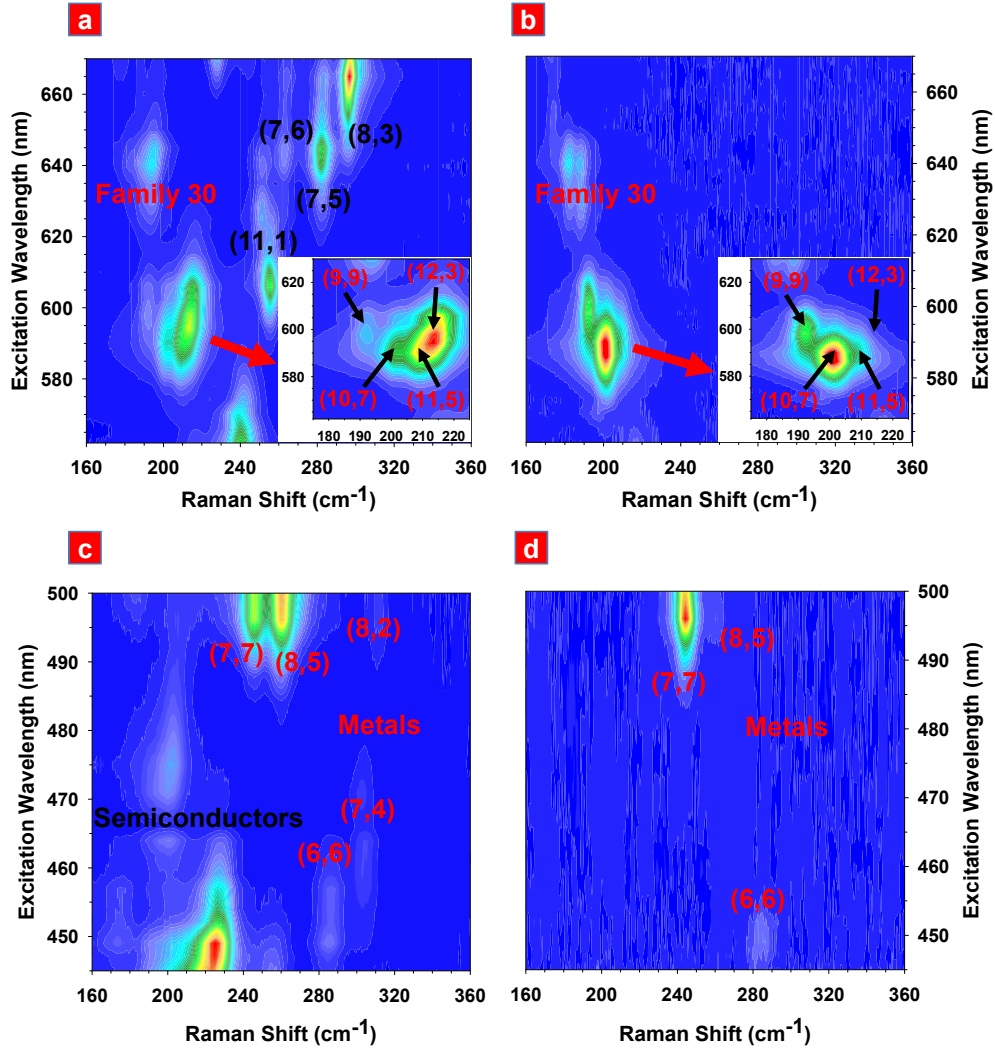

**Figure 2**. Resonant Raman scattering excitation maps demonstrating enrichment of armchair and near-armchair carbon nanotubes. (a,b) Resonant Raman scattering excitation maps of as-produced (a) and metallic-enriched (b) SWNT samples over an excitation range of 562-670 nm. The insets of a and b highlight family $2n + m = 27$. The scale of the inset of a has been magnified by 1.5 relative to the full excitation map to differentiate the (*n,m*) members of family 27 more clearly. (c,d) Resonant Raman scattering excitation maps of as-produced (c) and metallic-enriched (d) SWNT samples over an excitation range of 445-500 nm. Metallic (*n,m*) species are labeled in red and semiconducting species are labeled in black. In both excitation ranges, a clear shift in relative Raman intensity occurs towards metallic (*n,m*) species of large chiral angle, namely the armchair and near-armchair species.

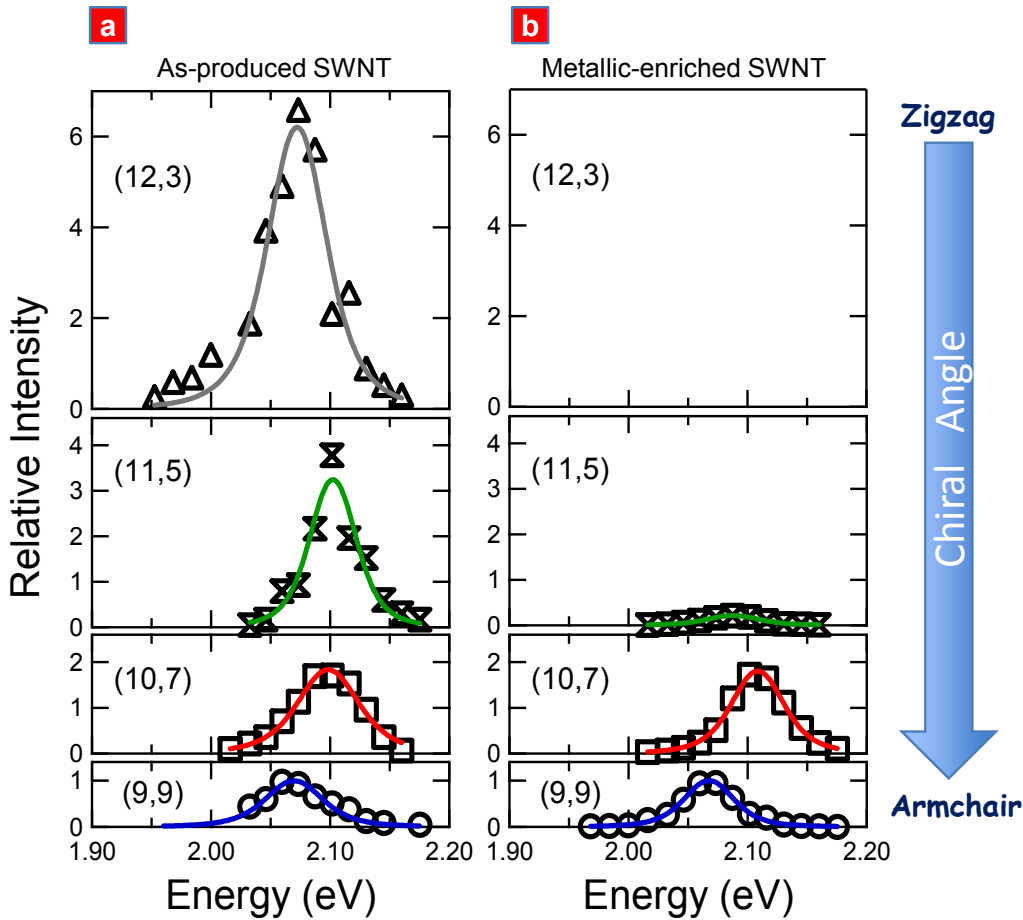

**Figure 3.** Resonant Raman excitation profiles for different (*n,m*) metallic nanotubes showing relative population change towards large chiral species via density gradient ultracentrifugation enrichment. (a) Raman excitation profiles (REPs) for the (n,m) members of family $2n + m = 27$ as observed in the as-produced SWNT sample. (b) Raman excitation profiles for the (n,m) members of family $2n + m = 27$ as observed in the metallic-enriched SWNT sample. The (12,3) is not displayed in b because an REP could not be constructed due to its extremely weak observation at only one excitation wavelength (586 nm). Near-zigzag species (12,3) and (11,5) see a marked decrease in intensity while armchair (9,9) and near-armchair (10,7) species remain undisturbed by the DGU process.

| (n,m) | Diameter [nm] | Chiral Angle [degree] | REP Amplitude of Family 2n+m (AP) | REP Amplitude of Family 2n+m (ME) | REP Electronic Broadening (AP) [meV] | REP Electronic Broadening (ME) [meV] | Theoretical Raman Intensity per unit length[a] | % Population of Family 2n+m (AP) | % Population of Family 2n+m (ME) |
|---|---|---|---|---|---|---|---|---|---|
| **Family 27** | | | | | | | | | |
| (9,9) | 1.221 | 30 | 1.0 | 1.0 | 80 | 67 | 0.02 | 36% | 52% |
| (10,7) | 1.159 | 24.2 | 2.5 | 2.3 | 87 | 71 | 0.05 | 34% | 46% |
| (11,5) | 1.111 | 17.8 | 2.2 | 0.3 | 77 | 71 | 0.13 | 11% | 2% |
| (12,3) | 1.077 | 10.9 | 6.6 | 0* | 81 | - | 0.24 | 18% | 0% |
| **Family 30** | | | | | | | | | |
| (10,10) | 1.357 | 30 | 1.0 | 1.0 | 90 | 83 | 0.01 | 40% | 46% |
| (11,8) | 1.294 | 24.8 | 1.2 | 1.8 | 72 | 76 | 0.03 | 16% | 27% |
| (12,6) | 1.244 | 19.1 | 5.2 | 5.5 | 95 | 109 | 0.09 | 22% | 27% |
| (13,4) | 1.206 | 13 | 8.0 | 0* | 91 | - | 0.16 | 19% | 0% |
| (14,2) | 1.183 | 6.6 | 2.5 | 0* | 66 | - | 0.24 | 4% | 0% |

**Table 1**. Evidence for enrichment of armchair and near-armchair carbon nanotubes. Tabulated data extracted from numerical fits to Raman excitation profiles of members of families $2n + m = 27$ and 30 for both the as-produced and metallic-enriched samples. Footnote: *a* The (*n,m*)-dependent theoretical Raman intensities per unit length were extracted from data displayed in Reference 22. * These species were only weakly observable at one or two excitation wavelengths near resonance and as such REPs could not be constructed for them.

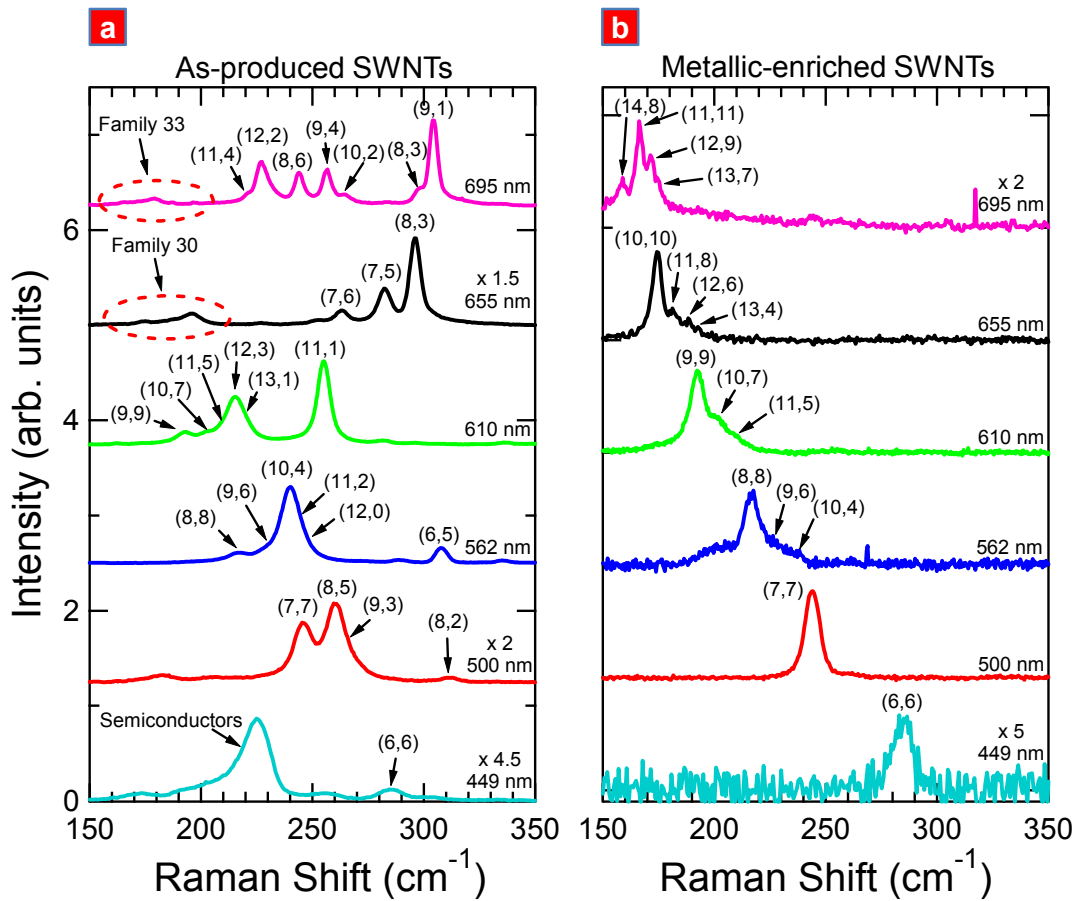

**Figure 4.** Suppression of semiconducting nanotubes and zigzag and near-zigzag metallic nanotubes. (a) Selected Raman spectra from data collected on the as-produced SWNT sample. (b) Selected Raman spectra from data collected on the metallic-enriched SWNT sample. These specific excitation wavelengths were chosen to highlight the armchair species. In (b), all semiconducting nanotubes are absent and small chiral angle metallic nanotubes are strongly suppressed, leaving only armchair and near-armchair nanotubes.

**Methods**

*Sample Preparation:*

Pristine, single-walled carbon nanotubes (SWNTs) were synthesized by the high-pressure carbon monoxide (HiPco) method at Rice University. The as-produced sample was produced by a variation of the standard ultracentrifugation technique[1]. HiPco SWNTs (batch HPR 188.1) were initially dispersed in 1% (wt./vol.) sodium deoxycholate (DOC) in water by bath sonication (Cole-Parmer 60 W ultrasonic cleaner, model #08849-00) for 30 minutes, using a starting SWNT concentration of 200 mg/L. The suspension was then further sonicated by probe ultrasonicator (Cole-Parmer 500 W ultrasonic processor, model # CPX-600, ¼" probe, 35% amplitude) for 30 minutes while being cooled in a water bath maintained at 10°C. Finally, the suspension was centrifuged for 4 hours at 109,500 g average (Sorvall Discovery 100SE Ultracentrifuge using a Sorvall AH-629 swing bucket rotor). After centrifugation, the upper 60% of the supernatant was removed and used for optical measurements.

Samples enriched in metallic SWNTs were produced by the density gradient ultracentrifugation (DGU) technique employing a three-surfactant system[2]. HiPco SWNTs (batch HPR 188.1) were initially dispersed in 1% (wt./vol.) DOC in water by bath sonication (Cole-Parmer 60W ultrasonic cleaner, model #08849-00) for 30 minutes. The starting concentration of SWNTs was 1 g/L. The suspension was then further sonicated by probe ultrasonicator (Cole-Parmer 500 W ultrasonic processor, model # CPX-600, ¼" probe, 35% amplitude) for 17.5 hours while being cooled in a water bath maintained at 10°C. The suspension was then centrifuged for 1 hour at 208,400 g average (Sorvall Discovery 100SE Ultracentrifuge using a Beckman SW-41 Ti swing bucket rotor) to remove large bundles of SWNTs. After centrifugation, the upper 80% of the supernatant was removed for use in DGU.

A mass density gradient was prepared composed of 1.5% (wt./vol.) sodium dodecyl sulfate (SDS), 1.5 % (wt./vol.) sodium cholate (SC), and varying amounts of iodixanol. The gradient was layered inside a centrifuge tube in 2 mL volume steps starting from the bottom with 40% (wt./vol.), 30% (wt./vol.), 27.5% (wt./vol.), 25% (wt./vol.), 22.5% (wt./vol.), and 20% (wt./vol.) iodixanol. All gradient steps except the 30% layer contained 1.5% (wt./vol.) SDS and 1.5% (wt./vol.) SC. The 30% layer, which contained SWNTs, was prepared by mixing 1 mL of the SWNT supernatant prepared previously with 1 mL of 60% iodixanol, 2% SDS, and 2% SC to ultimately form 2 mL of 30% iodixanol, 1% SDS, 1% SC, and 0.5% DOC.  The gradient was then centrifuged for 18 hours at 208,400 g average (Beckman SW-41 Ti swing bucket rotor).  The resulting separated material was then removed by hand pipetting in 200 μL fractions with the most metallically enriched material appearing at the top of a resulting pink band.  The collected fractions were then dialyzed into a 1% DOC (water) solution (Pierce, 3500D MW dialysis cassette) and used for optical measurements.

*Optical Measurements:*

Optical absorption spectroscopy was performed in the 190-1400 nm range in 1 nm steps on an ultraviolet-visible-near-infrared, double beam spectrophotometer (Shimadzu UV-3101PC scanning spectrophotometer) through a 10 mm path length quartz cuvette using a 1% (wt./vol.) DOC (water) reference.

Photoluminescence excitation spectroscopy was performed using excitation light with a 5 nm bandwidth in the wavelength range of 500-850 nm, obtained from a Xe lamp using a double monochromator (HORIBA Jobin-Yvon Fluorolog-3-211).  Nanotube emission was measured from 850 to 1400 nm with a single-channel, cooled InGaAs detector via a monochromator with 6 nm bandpass filter.  Spectra were acquired with 2.5 nm and 4.0 nm steps in excitation and emission wavelengths, respectively.  Individual spectra were acquired with a 1 sec per point integration time.  Acquisitions were repeated, as required, to ensure reproducibility and to improve signal to noise ratio.  Spectra were corrected for power and instrument response.

Resonant Raman spectroscopy was performed in a backscattering configuration with cw Ti:Sapphire laser excitation, tunable dye laser excitation using Kiton Red and Rhodamine 6G dyes, $Ar^+$ ion laser discrete lines and frequency-doubled cw Ti:Sapphire laser excitation scanned from 850-695 nm, 680-610 nm, 615-562 nm, 514.5 and 501.7 nm, and 500-440 nm, respectively.  Excitation power was maintained at 25 mW. Individual Stokes-shift spectra were obtained as 5 min integrations using a charge coupled device camera mounted on a SPEX triple monochromator. The frequency of each carbon nanotube spectrum was calibrated at each excitation wavelength with the non-resonant Raman spectrum of 4-acetamidophenol. Intensities were corrected for instrument response using fits to the intensities of peaks of 4-acetamidophenol and scaling the nanotube radial breathing mode spectra by the average intensity value at each excitation wavelength. All Raman spectra were taken at room temperature and ambient pressure.

**Residual Semiconducting ($n,m$) Species**

*Resonant Raman Scattering Data:*

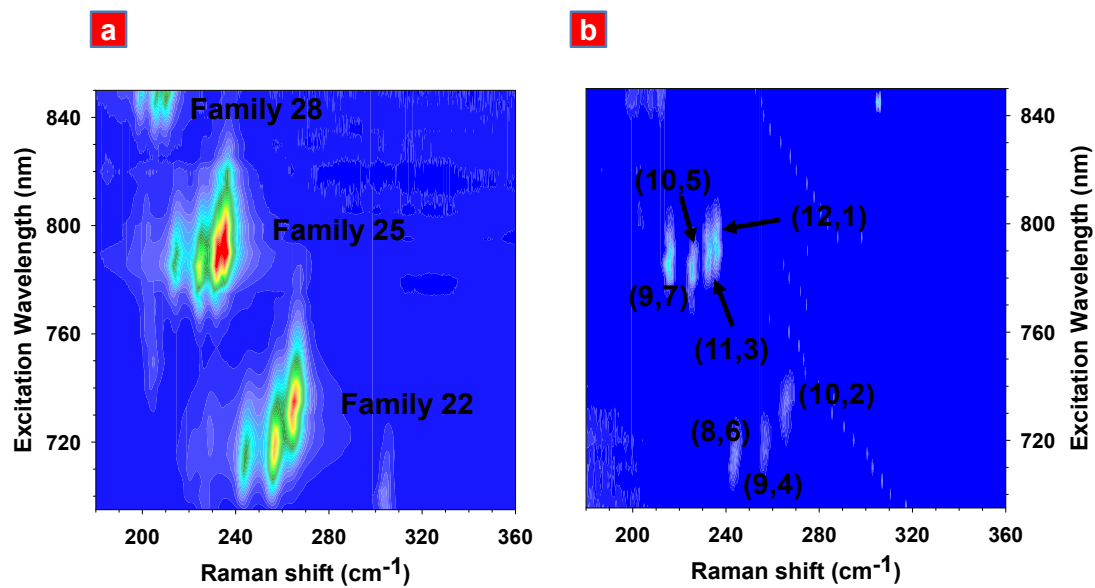

**Figure S1.** Suppression of semiconducting nanotubes. (a,b) Resonant Raman scattering excitation maps of the radial breathing modes of as-produced (a) and metallic-enriched (b) SWNT samples over an excitation range of 695-850 nm. A strong and clear suppression of semiconducting ($n,m$) species is observed for members of ($2n + m$) families = 22, 25, and 28.

Supplementary Figs. S1a and S1b show RRS spectra plotted as a two-dimensional contour plot as a function of excitation wavelength (695-850 nm) and Raman shift of the radial breathing mode (RBM) for AP-SWNT and ME-SWNT samples, respectively. In this excitation range for HiPco nanotubes, the optical response primarily results from semiconducting species in the 0.75-1.20 nm diameter range excited via $E_{22}^S$. Immediately, the strong suppression of RBMs assigned to semiconductors of the ($2n + m$) families 22, 25, and 28[3] is observed in Fig. S1**b** (ME-SWNT) although the suppression is not complete, indicating that some trace amounts of these ($n,m$)-species remain. Excellent agreement between the peak positions of Fig. S1**b** and those of the individualized semiconductors present in Fig. S1**a** (AP-SWNT) indicates that the remaining semiconducting species present are as individual nanotubes and not as parts of large bundles[3-4].

*Photoluminescence Data:*

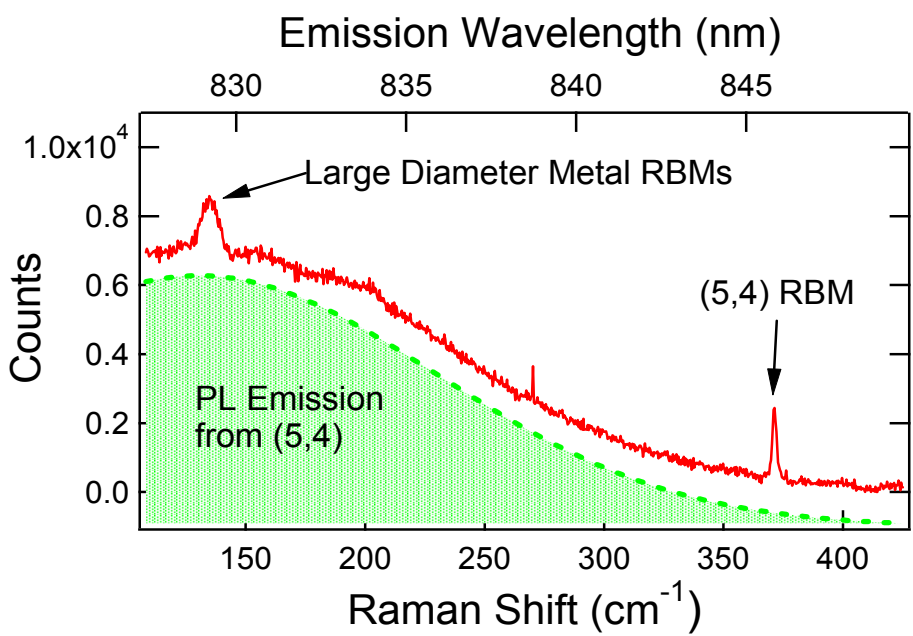

**Figure S2.** The presence of weak photoluminescence in ME-SWNT material. A Raman spectrum excited at 820 nm shows the appearance of a weak semiconducting (5,4) nanotube RBM on top of a background due to weak PL emission also from (5,4). The PL emission has been offset from the raw Raman data for clarity.

Referring back to our earlier discussion of PLE data and in light of the previous statements above, we can confidently discard the alternative explanation for lack of emission from the ME-SWNT sample mentioned and accept the original conclusion that the lack of emission is a result of trace amounts of semiconducting nanotubes below the detection threshold of the instrument. The apparent discrepancy in detection of semiconductors in RRS versus PLE is resolved when one examines the experimental conditions used for both measurements (see above Supplementary Information - *Optical Measurements*) and one observes that PLE employed a substantially weaker excitation source (~1 mW from a Xe lamp was employed for PLE measurements versus a laser at 25 mW for RRS measurements) and shorter acquisition time (1 sec for PLE versus 5 min for RRS) than for RRS. In fact, close examination of raw, uncorrected Raman spectra in ME-SWNT reveal very weak emission from trace semiconductors as is shown in Fig. S2 taken with 820 nm excitation. Here, PL emission from the (5,4) is detected as part of the background of the spectrum, clearly showing the (5,4) RBM excited via $E_{11}^S$.

**Suppression of (9,3) Member of Family (2*n* + *m*) = 21**

Figure S3 shows Raman spectra excited using a discrete line from an $Ar^+$ ion laser at 514.5 nm. While the AP-SWNT material (top, red trace) shows the members of family (2*n* + *m*) = 21, including the (9,3), and one member of family (2*n* + *m*) = 18, the ME-SWNT sample (bottom, blue trace) clearly shows the complete suppression of the (8,5) and (9,3) from the (7,7) family and the (8,2) from the (6,6) family, leaving behind only the (7,7) armchair.

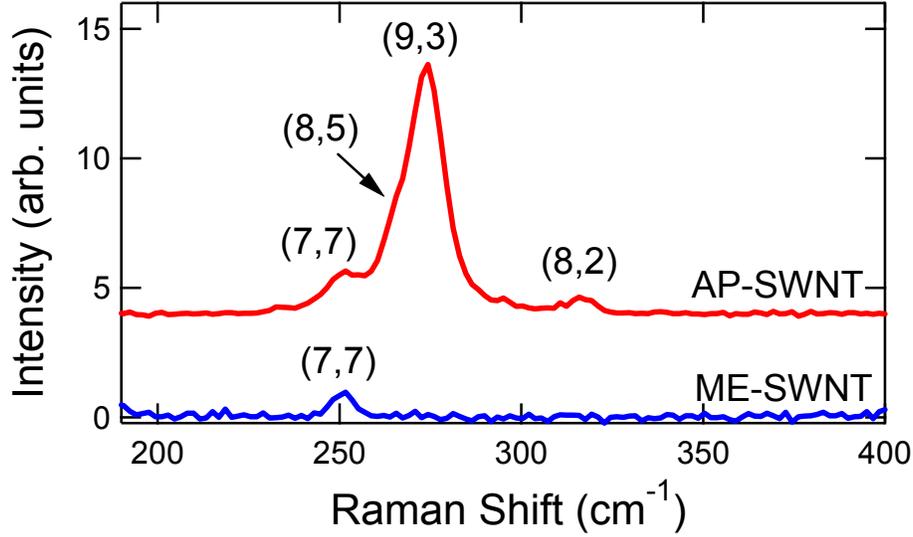

**Figure S3.** Suppression of (9,3) member of family $(2n + m) = 21$. Raman spectra excited using a discrete line from an Ar$^+$ ion laser at 514.5 nm. While the AP-SWNT material (top, red) shows strong signal from non-armchair members of family $(2n + m) = 21$, the ME-SWNT sample (bottom, blue) clearly shows the complete suppression of the (8,5) and (9,3) from the (7,7) family and the (8,2) from the (6,6) family, leaving behind only the (7,7) armchair. The AP-SWNT trace is offset for clarity.

**REP Fitting and Table Calculations:**

Raman excitation profiles were constructed from data presented in Figs. 2a and 2b. For a particular $(n,m)$ (i.e., RBM phonon frequency), Raman intensity was plotted as a function of excitation photon energy (in eV). Each REP was then fit, as an independent function of excitation photon energy and fixed phonon energy and the dependent parameters of optical transition energy, electronic broadening factor, and REP fitting amplitude, to a modified form of Equation (1) from the main text, resulting in the fitting equation (S1):

$$I_{Raman}^{Fit} = \frac{A_{\exp}}{\left([E_{laser} - E_{ii}]^2 + \frac{1}{4}\gamma^2\right)\left([E_{laser} - E_{ii} - \hbar\omega_{ph}]^2 + \frac{1}{4}\gamma^2\right)} \quad (S1)$$

where $E_{laser}$, $E_{ii}$, $\gamma$, and $\hbar\omega_{ph}$ are defined previously, and $A_{exp}$ is the REP fitting amplitude $A_{exp} = N \cdot M_{e-o}^4 M_{e-ph}^2$ where $N$, $M_{e-ph}$, and $M_{e-o}$ are defined as before. Since the spectra were all corrected for instrumental response, the experimental prefactor $g$ is removed from our fitting equation. From these values, enrichment of armchair and near-armchairs is already evident by the approximately constant ratio of REP amplitude between armchair and near-armchair species between AP-SWNT and ME-SWNT samples [i.e., $A_{exp}(10,7)/A_{exp}(9,9) = 2.5$ and 2.3 for AP- and ME-SWNT samples, respectively] as opposed to the large decrease in REP amplitude ratio between armchair and near-zigzag species [i.e., $A_{exp}(11,5)/A_{exp}(9,9) = 2.2$ and 0.3 for AP- and ME-SWNT samples, respectively].

To further quantify this change in relative REP amplitude into a change in relative population of each $(n,m)$ species[5] within a $(2n + m)$ family, one must correct the Raman intensity for the chiral angle and diameter dependences of the exciton-phonon and exciton-photon coupling matrix elements. Using the theory developed by Jiang et al.[6], where the electronic broadening factor, $\gamma_{theory}$, is set to be 60 meV for all $(n,m)$ species considered, we extracted values for the Raman intensity per unit length (essentially removing the $N$-dependence from Raman intensity) at the resonance condition for each $(n,m)$ REP measured in our experiments. As such, Equation (S1) reduces to:

$$I_{Raman}^{E_{laser}=E_{ii}} = \frac{A_{theory}}{\left(\hbar^2\omega_{ph}^2 + \frac{1}{4}\gamma_{theory}^2\right)\left(\frac{1}{4}\gamma_{theory}^2\right)} \quad (S2)$$

for the theoretical Raman intensity per unit length, where $A_{theory} = M_{e-o}^4 M_{e-ph}^2$ and $\gamma_{theory} = 60$ meV. We can then rearrange Equation (S2) and calculate $A_{theory}$ as a function of the theoretical Raman intensity per unit length. Using that, we obtain an expression for the relative population of each $(n,m)$ as

$$N \propto \frac{A_{exp}}{A_{theory}} \quad (S3).$$

The relative populations can then be converted to percent populations of each $(2n + m)$ family [in this case $(2n + m)$ families = 27 and 30] by dividing the relative population for a particular $(n,m)$ by the sum of relative populations of all the $(n,m)$ members of that particular family. From this viewpoint, we can see that the armchair (9,9), which made up 36% of family 27 in the AP-SWNT sample now makes up 52% in the ME-SWNT sample.

**Supplementary references**

1. O'Connell, M. J.; Bachilo, S. M.; Huffman, C. B.; Moore, V. M.; Strano, M. S.; Haroz, E. H.; Rialon, K. L.; Boul, P. J.; Noon, W. H.; Kittrell, C.; Ma, J.; Hauge, R. H.; Weisman, R. B.; Smalley, R. E. *Science* **2002**, *297*, 593-596.

2. Yanagi, K.; Miyata, Y.; Kataura, H. *Appl. Phys. Exp.* **2008**, *1*, 034003.

3. Doorn, S. K.; Heller, D. A.; Barone, P. W.; Ursey, M. L.; Strano, M. S. *Appl. Phys. A* **2004**, *78*, 1147-1155.

4. O'Connell, M. J., Sivaram, S. & Doorn, S. K. *Phys. Rev. B* **2004**, 69, 235415.

5. Jorio, A.; Santos, A. P.; Ribeiro, H. B., Fantini, C.; Souza, M.; Vieira, J. P. M.; Furtado, C. A.; Jiang, J.; Saito, R.; Balzano, L.; Resasco, D. E.; Pimenta, M. A. *Phys. Rev. B* **2005**, 72, 075207.

6. Jiang, J.; Saito, R.; Sato, K.; Park, J. S.; Samsonidze, Ge. G.; Jorio, A.; Dresselhaus, G.; Dresselhaus, M. S. *Phys. Rev. B* **2007**, *75*, 035405.